\documentclass[aps,prc,fleqn,floatfix,twocolumn,twoside]{revtex4}

\usepackage[dvipdfm]{graphicx}% Include figure files

\usepackage{amsmath,amssymb,amsfonts}

\usepackage{color}

\begin{document}

\title{Counting hot/cold spots in quark-gluon plasma}

\author{Guang-You Qin and Berndt M\"uller}
\affiliation{Department of Physics, Duke University, Durham, NC 27708, USA}$$$$

\date{\today}
%%%%%%%%%%%%%%%%%%%%%%%%%%%%%%%%%%%%%%%%%%%%%%%%%%%%%%%%%%%%%%%%%%%%%%%%%%%%%%%%
%%%%%%%%%%%%%%%%%%%%%%%%%%%%%%%%%%%%%%%%
\begin{abstract}

We study how local fluctuations in the initial states of relativistic heavy-ion collisions manifest themselves in the correlations between different orders of harmonic moments of the density profiles, particularly those involving only odd harmonics which purely arise from initial state fluctuations.
We find the strengths of those correlations are sensitive to the number of hot and cold spots in the initial states.
Hydrodynamic evolution of the fireball translates initial state geometric anisotropies as well as their correlations into final state momentum anisotropies and correlations.
We conclude that the measurement of the correlations between different harmonic moments of final state azimuthal distribution can be employed to quantify the inhomogeneity of the initial density profiles such as the population of hot and cold spots that are produced in high energy nuclear collisions.

\end{abstract}
\maketitle
%%%%%%%%%%%%%%%%%%%%%%%%%%%%%%%%%%%%%%%%%%%%%%%%%%%%%%%%%%%%%%%%%%%%%%%%%%%%%%%%
%%%%%%%%%%%%%%%%%%%%%%%%%%%%%%%%%%%%%%%%

The creation of hot, dense quark-gluon plasma (QGP) can be achieved by colliding two heavy nuclei at ultra-relativistic energies, such as those at the Relativistic Heavy-Ion Collider (RHIC) and the Large Hadron Collider (LHC).
The dynamical evolution of the QGP produced in these energetic collisions, particularly the large collective flow observed at RHIC and the LHC \cite{Adams:2003zg, Aamodt:2010pa, Collaboration:2011yk}, can be well simulated by relativistic hydrodynamics \cite{Kolb:2003dz}.
Being the hydrodynamic response to the pressure gradients, the collective flow exhibited by such highly exited QCD matter is azimuthally anisotropic in the plane transverse to the beam axis in the collisions with nonzero impact parameter.
The anisotropy of the flow is usually quantified by the Fourier expansion coefficients $v_n$ of the final state momentum distribution in the transverse plane \cite{Ollitrault:1992bk}.
There has been extensive study of elliptic flow $v_2$ as a function of various
quantities, mostly aiming for a quantitative extraction of the transport properties, e.g., the shear viscosity to entropy ratio, of the produced hot QCD matter \cite{Luzum:2008cw, Dusling:2007gi, Schenke:2010rr, Song:2010mg}.

Early comparisons to the measured elliptic flow $v_2$ using relativistic hydrodynamic simulation employed a set of smooth, event-averaged initial conditions which are specified by the event-averaged geometry of the collision zone, usually in terms of energy or entropy density at initial times.
However, the outcome from each nucleus-nucleus collision fluctuates event-by-event due to quantum fluctuations, such as the positions of nucleons or color charges inside the colliding nuclei \cite{Miller:2003kd, Broniowski:2007ft, Alver:2008zza, Hirano:2009bd}.
One consequence of such fluctuations in the initial states is the finite eccentricity and elliptic flow even in the collisions with almost zero impact parameter \cite{Alver:2006wh}.
Another significant feature of the initial state fluctuations is the presence of odd harmonic moments in the initial geometric anisotropy and final momentum anisotropy \cite{Alver:2010gr, Alver:2010dn, Petersen:2010cw, Shuryak:2009cy, Staig:2010pn, Qin:2010pf, Lacey:2010hw, Nagle:2010zk, Ma:2010dv, Xu:2010du, Teaney:2010vd, Qiu:2011iv, Bhalerao:2011yg, Floerchinger:2011qf}, in addition to even harmonic moments which are just the reflection of the symmetry of the overlap region between two colliding nuclei at finite impact parameters.
The fluctuating non-smooth initial conditions have been invoked to explanin the double-hump structure and near-side ridge phenomenon in two-particle correlation measurements \cite{Abelev:2008un, Alver:2009id, Sorensen:2008dm, Takahashi:2009na, Andrade:2009em, Alver:2010gr, Sorensen:2010zq}.
The measurements of azimuthal anisotropy of the final momentum distribution from RHIC and the LHC have shown a prominent third harmonic flow $v_3$ and other odd harmonic flows, in addition to even harmonic flow coefficients \cite{Adare:2011tg, :2011vk, Collaboration:2011hf}.
Significant attention is now devoted to analyzing the dynamical evolution of initial state fluctuations and their hydrodynamic responses and translation to final harmonic flows \cite{Petersen:2010cw, Qin:2010pf, Schenke:2010rr, Teaney:2010vd, Qiu:2011iv}. The goal is to achieve a quantitative understanding of the expansion dynamics of the fireball produced in these energetic nucleus-nucleus collisions.

In this work, we study another important aspect of initial state fluctuations: the inhomogeneity of the initial density profile.
The local fluctuations (hot or cold spots) not only generate geometric anisotropy in terms of harmonic moments and their fluctuations, different orders of harmonic moments are actually correlated with each other. Some of this type of correlations have been investigated in earlier studies \cite{Staig:2010pn, Teaney:2010vd, Bhalerao:2011yg}.
Our objective is to study how the local fluctuations on top of event-averaged profiles manifest in the correlations between different harmonic moments of the initial state geometric anisotropy.
Due to hydrodynamic evolution of the fireball, these correlations will be translated to the correlations between final state particles.
Our particular interests are focused on those correlations involving only odd harmonic moments which purely stem from the local fluctuations in the initial states.
We show that the correlations between different harmonic moments can be employed to study the degree of inhomogeneity of the initial density profile. Especially, the magnitudes of the correlations strongly depend on the number of hot/cold spots present in the initial states.
Once measured by experiments, these correlations may infer much information about the initial state fluctuations, such as the population of hot/cold spots inside the hot, dense QGP produced in ultra-relativistic nuclear collisions.

For a given initial energy or entropy density profile $f(\vec{r}) = f(r,\phi)$ in the transverse plane, we may quantify the anisotropy of the profile by defining eccentricities $\epsilon_n$,
\begin{eqnarray}
\epsilon_n = \{r^m \cos(n\phi - n\Phi_n)\} / \{r^m\}
\label{epsn}
\end{eqnarray}
Here we use curly brackets $\{ \cdots \}$ to represent the average of density profile within a given event and angle brackets $\langle \cdots \rangle$ for the average of quantities over many events (the ensemble average).
The initial spatial event plane angle $\Phi_n$ for the $n$-th harmonic moment is define as
\begin{eqnarray}
\Phi_n = \frac{1}{n} \arctan\left[\{r^m \sin(n\phi)\} / \{r^m \cos(n\phi)\}\right]
\end{eqnarray}
where $\phi = \arctan(y/x)$ represent the polar angle for the point $(x,y)$.
Here we take the exponent in $r^m$ to be $m=n$ for $n \ge 2$ \cite{Qin:2010pf} and $m=3$ for the dipole asymmetry $n=1$ following Teaney and Yan \cite{Teaney:2010vd}, who introduced a cumulant expansion to parameterize the initial conditions. Gardim {\em et al.} \cite{Gardim:2011xv} have argued that the choice adopted here serves a better estimator than $m=2$ for the flow harmonics $v_n$ with $n \ge 2$.

Being the hydrodynamic response to the initial state spatial anisotropy $\epsilon_n$, the momentum anisotropy parameters $v_n$ (flow harmonics) of the final state particle azimuthal angle ($\psi$) distribution are defined as
\begin{eqnarray}
v_n = \{\cos(n\psi - n\Psi_n) \}
\end{eqnarray}
where $\Psi_n$ is the final (momentum) event plane angle.
In an event-by-event hydrodynamic analysis with fluctuating initial conditions, the final momentum event plane $\Psi_n$ is strongly correlated to the initial spatial event plane $\Phi_n$ \cite{Holopainen:2010gz, Petersen:2010cw, Qiu:2011iv}.
Assuming linear hydrodynamic responses of anisotropic flows $v_n$ to the initial spatial anisotropy $\epsilon_n$, one may set $\Psi_n = \Phi_n + \pi/n$.
Note there may exist non-linear hydrodynamic response of $v_n$ to $\epsilon_n$, e.g., $v_4$ can be developed from $\epsilon_4$ as well as $\epsilon_2$.

In relativistic nuclear collisions, the initial states fluctuate from one event to another due to local fluctuations.
Thus for a given event we may write down the initial density profile $f(\vec{r})$ in the transverse plane as the sum of an event-averaged profile $f_0(\vec{r})$ and an event-by-event fluctuating one $\delta f(\vec{r})$,
\begin{eqnarray}
f(\vec{r}) = f_0(\vec{r}) + \delta f(\vec{r})
\end{eqnarray}
By definition, $\langle f(\vec{r})\rangle = f_0(\vec{r})$ and $\langle \delta f(\vec{r}) \rangle = 0$.
The event-averaged profile $f_0(\vec{r})$ is usually specified in a particular model; here we use the optical Glauber model \cite{Glauber:1970jm, Miller:2007ri}, with the initial density being proportional to the combination of binary collision density and participant nucleon density \cite{Kharzeev:2000ph}: $f_0(\vec{r}) \propto \left[\alpha n_{\rm coll}(\vec{r}) + (1-\alpha) n_{\rm part}(\vec{r})/2 \right]$.
For the simulation of the collisions of two lead-lead collisions at $\sqrt{s_{NN}} = 2.76$~TeV at the LHC, we use Woods-Saxon profile for the the nuclear density, with the radius parameter taken to be $R_{\rm ws} = 6.62$~fm and the diffusion parameter $d=0.546$~fm. The inelastic nucleon-nucleon cross section is taken to be $\sigma_{NN} = 64$~mb.
The parameter $\alpha$ is chosen as $0.11$ for a good description of the centrality-dependence of the charged hadron multiplicity distribution \cite{Aamodt:2010cz}.

The harmonic moments $\epsilon_n$ and $\Phi_n$ can be expressed using event-averaged and fluctuating profiles. They are particularly simple for odd harmonic moments:
\begin{eqnarray}
\Phi_n \!\!&=&\!\! \frac{1}{n} \arctan\left[{\{r^m \sin(n\phi)\}_\delta} / {\{r^m \cos(n\phi)\}_\delta}\right]
\nonumber \\
\epsilon_n \!\!&=&\!\!  \{r^m \cos(n\phi - n\Phi_n)\}_\delta / (\{r^m\}_0 + \{r^m\}_\delta)
\label{epsn_odd}
\end{eqnarray}
where the subscripts in $\{\cdots\}_0$ and $\{\cdots\}_\delta$ represent the average over the profiles $f_0(\vec{r})$ and $\delta f(\vec{r})$, respectively.
We see that for odd harmonic moments, the event-plane angles $\Phi_n$ are completely determined by the fluctuating part of the profile $\delta f(\vec{r})$, and the event-averaged profile $f_0(\vec{r})$ only contributes to the overall normalization for the eccentricities $\epsilon_n$.

For given event-by-event initial conditions including various fluctuations, we may obtain the event distribution of the fluctuating part of the profile $\delta f(\vec{r})$. In this work, we model the fluctuating part $\delta f(\vec{r})$ for a given event by a combination of $N$ local fluctuations,
\begin{eqnarray}
\delta f(\vec{r}) = \sum_{i=1}^N \frac{s_i A_h(s_i)}{\pi R_h^2} \theta\left(R_h-|\vec{r}-\vec{r}_i|\right)
\end{eqnarray}
Here, $\vec{r}_i = (r_i,\phi_i)$ is the location of the fluctuation with radius of $R_h$ (taken to be the same for all fluctuations).
The sign factor $s_i=\pm 1$ represents whether the local fluctuation is positive or negative (hot or cold spot) and $A_h(s_i)$ represents the amplitude of the fluctuation.
Both hot and cold spots are generally called hot spots in what follows.
The local fluctuations are assumed to be small perturbations on top of the event-averaged profile, i.e., $A_h(\pm 1)N \ll \int d^2\vec{r} f_0(\vec{r})$.
The probabilities for hot and cold spots are related to their amplitudes by $P_{s_i}(s_i=1) A_h(1) = P_{s_i}(s_i=-1) A_h(-1)$ to ensure $\langle \delta f(\vec{r}) \rangle = 0$.

The description of initial state fluctuations with hot and cold spots is the approximation of the realistic event distribution $P(\delta f)$ by two spikes located at $A_h(1)/(\pi R_h^2)$ and $-A_h(-1)/(\pi R_h^2)$ with the heights being $P_{s_i}(1)$ and $P_{s_i}(-1)$, in addition to a spike at zero (whose height, denoted as $P_{s_i}(0)$, is not relevant to the following discussion).
The heights of two spikes are usually not equal since the event distribution $P(\delta f)$ in realistic initial conditions is not symmetric around zero.

As mentioned earlier, the presence of hot spots generate not only the spatial anisotropy in terms of harmonic moments $\epsilon_n$, but the correlations between different orders of harmonic moments as the hot spots are coherent combinations of all orders of harmonic moments.
In particular, the event plane angles $\Phi_n$ of different orders of harmonics are correlated to each other.
Such correlations can be quantified by the functions $C^{(k)}(n_1,n_2\cdots n_k)$,
\begin{eqnarray}
\label{Cmn_epsn}
C^{(k)} = \langle \cos[n_1\Phi_{n_1} + n_2\Phi_{n_2} + \cdots + n_k\Phi_{n_k}] \rangle \ \ \
\end{eqnarray}
Here we are interested in the correlation functions with the combinations of indices being $n_1+n_2+ \cdots + n_k = 0$ as these are not dependent on the particular choices of the coordinate.
For relativistic heavy-ion collisions, the above correlation functions are completely determined by the early time dynamics of the collisions.
Hydrodynamic evolution of the fireball will translate these correlations into the correlations in the final states.
With the substitution of $\Psi_n=\Phi_n + \pi/n$, we may cast the above correlation functions in terms of final momentum event plane angles,
\begin{eqnarray}
\label{Cmn_vn}
C^{(k)} = (-1)^k \langle \cos[n_1\Psi_{n_1} + n_2\Psi_{n_2} + \cdots + n_k\Psi_{n_k}] \rangle  \ \ \
\end{eqnarray}
This simplified substitution will be modified by the nonlinearity of the hydrodynamic equations, however, we expect that any such correlations in the initial state will survive the hydrodynamic evolution and manifest themselves in the final state, especially when the fluctuations have small amplitude.
Of particular interests are these involving only odd harmonic moments as they are purely from initial state fluctuations and less affected by the collision geometry.
For illustration purpose, here we consider these combinations involving only the first three odd harmonic moments, $C^{(4)}(1,1,1,-3)$, $C^{(4)}(1,1,3,-5)$ and $C^{(4)}(3,3,-1,-5)$.

To investigate how the hot spots on top of event-averaged profile affect the correlations between different harmonic moments, we need to model the spatial distribution of hot spots in the initial states.
Here we take two typical distributions of hot spots for comparison purpose.
The first one is the hard sphere model in which hot spots are sampled according to a uniform distribution in the overlap region of two hard sphere nuclei: $P_{HS}(\vec{r}) \propto \theta(R_{HS} - |\vec{r}-\vec{b}/2|) \theta(R_{HS} - |\vec{r}+\vec{b}/2|)$.
The other is taken as the two-component Glauber model in which hot spots are more distributed towards the center compared to the hard sphere model.

\begin{figure}[thb]
\includegraphics[width=0.9\linewidth]{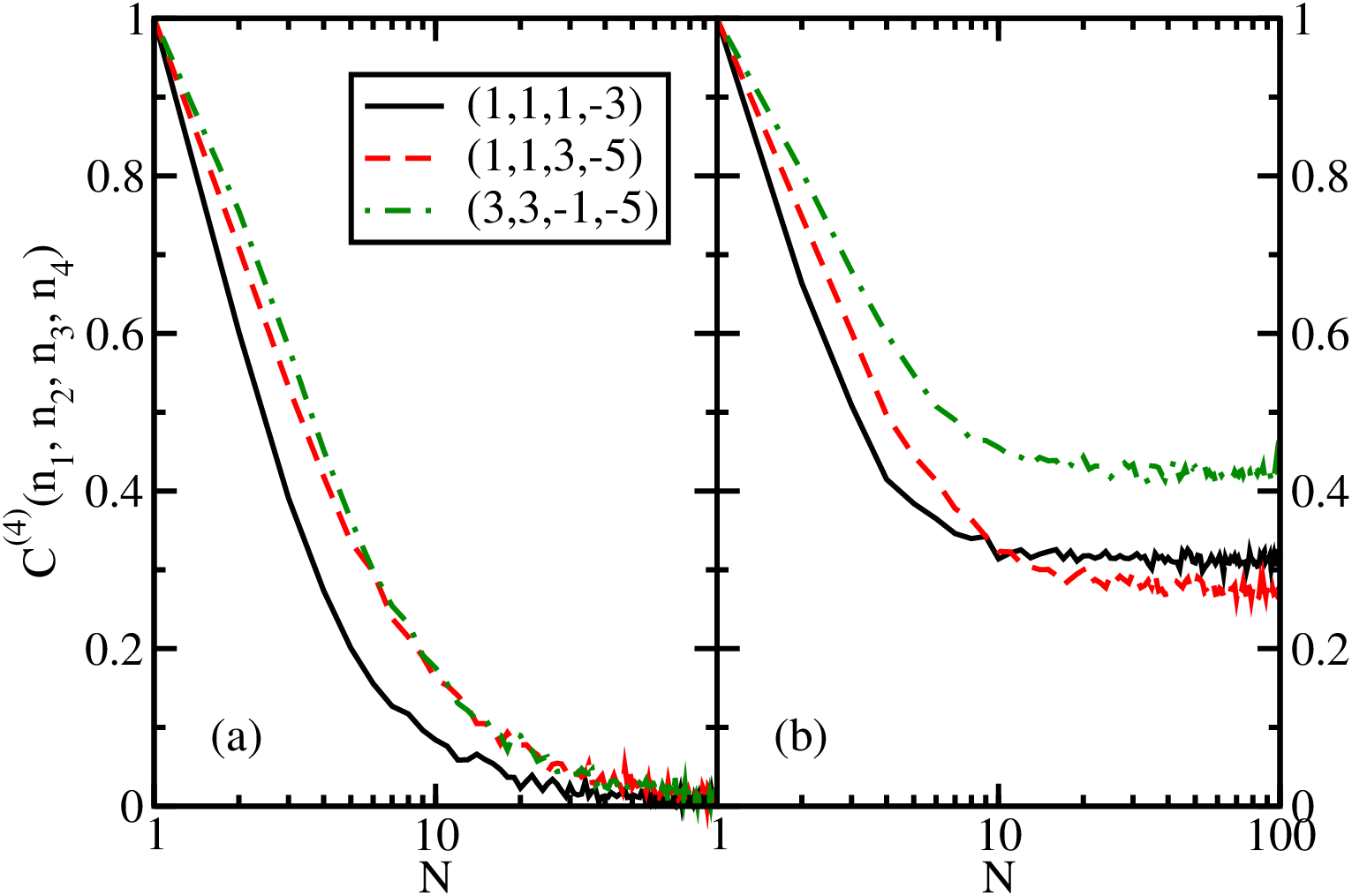}
\includegraphics[width=0.9\linewidth]{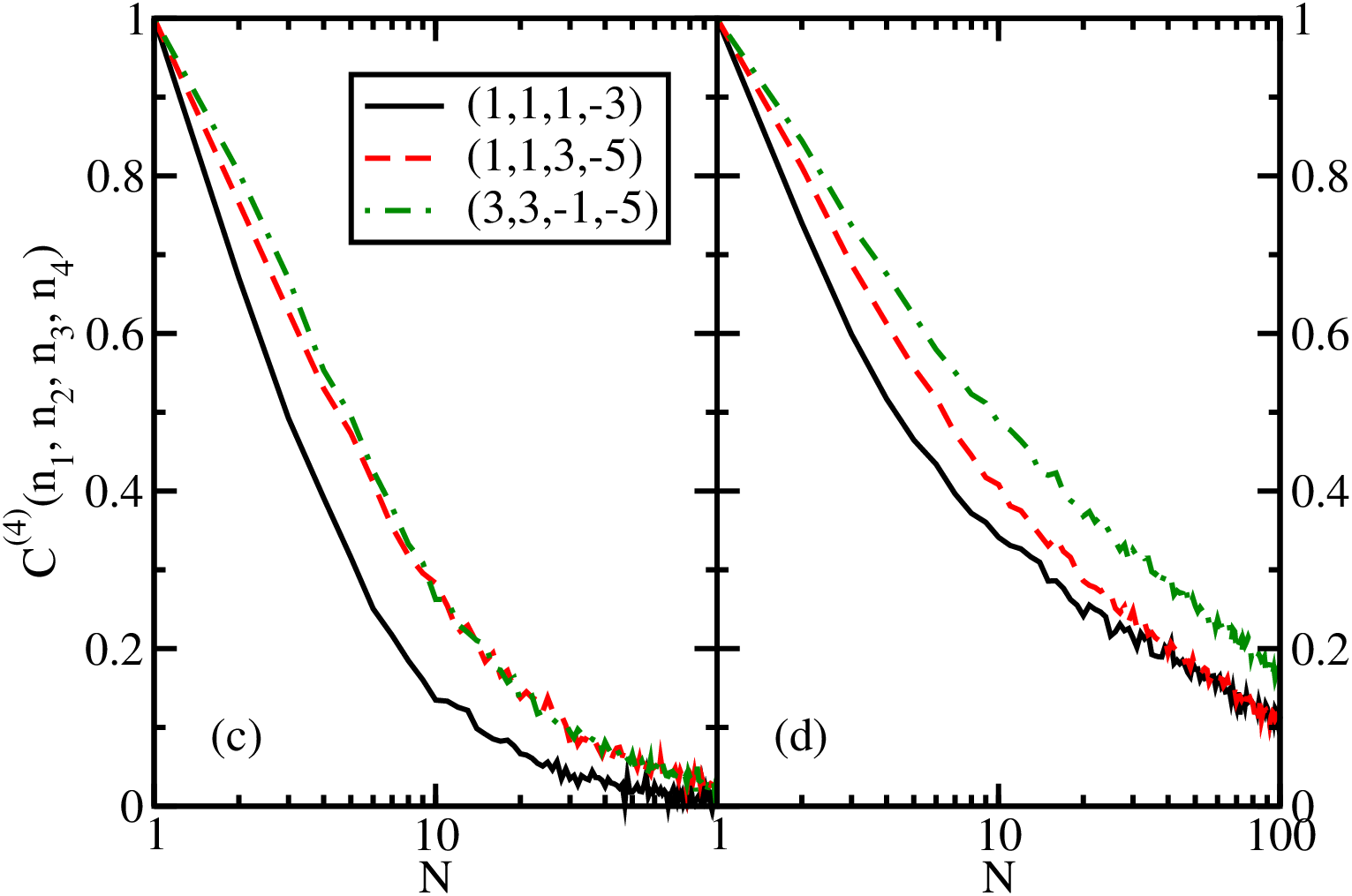}
 \caption{(Color online) The correlation functions $C^{(4)}$ as a function of the number of hot spots in the initial states. The panels (a, b) are for the hard sphere and (c, d) for the Glauber model; the panels (a, c) are for isotropic and (b, d) for anisotropic event distributions of hot spots.
} \label{cosmn}
\end{figure}

In Fig.~\ref{cosmn}, we show the event-averaged correlation functions $C^{(4)}$ as a function of the number of hot spots in the initial states.
The panels (a, b) show the results for the hard-sphere model of hot spot distribution, and (c, d) for the Glauber model.
The panels (a, c) show the results for the isotropic event distribution of hot spots in the transverse plane ($b=0$), while (b, d) for anisotropic distribution ($b\neq 0$).
Here for anisotropic hot spot distributions, we choose the impact parameter $b=12$~fm between two colliding nuclei (hard spheres or Woods-Saxon profiles); the hard sphere radius $R_{HS}$ is tuned to give similar values of eccentricity $\epsilon_2$ as the Glauber model when calculated with the event distributions of hot spots.
The radius $R_h$ of the hot spots is taken to be $0.25$~fm.
The increase of hot spot radius tends to decrease the correlations between different harmonic moments since hot spots are more evenly distributed in a given event.
Such effect is small for large system size or small number of hot spots.
The relative probabilities of hot and cold spots are taken as $P_{s_i}(1) = 35\%$ and $P_{s_i}(-1) = 65\%$ by comparing to the values of the correlation functions $C^{(3)}$, e.g., $C^{(3)}(2, 3, -5) \approx 0.1$ from Monte-Carlo Glauber modeling of lead-lead collisions at $b=0$~fm and our model with about a hundred hot spots.

The effects of the hot spots on these correlation functions are clearly seen from these plots.
If there is only one hot spot in the initial state, all correlation functions are equal to unity since all event plane angles are aligned or anti-aligned to each other, the strongest correlation that can be achieved.
With increasing number of hot spots, the interference between different hot spots decreases the correlations between different harmonic moments.
The geometric effect on the correlation functions can be seen by comparing (a) and (b) [or (c) and (d)] panels.
When the initial state contains few hot spots in, such effect is small since the hot spots are barely correlated with the geometry in each event.
But when the number of hot spots is large, all the event plane angles are strongly biased by the geometry in each event, thereby enhancing the correlations between different harmonic moments.

Due to different spatial ($r$) distribution of hot spots, there exists some difference in the correlation functions between these two models. In central collisions, the correlations are slightly stronger in the Glauber model distribution of hot spots. This is due to the fact that the hot spots are more distributed toward the center, thus fewer hot spots lie close to the edge of the overlap region.
This also reduces the dependence on the geometry of the hot spot distribution, i.e., smaller enhancement of the correlations from isotropic distribution of hot spots to anisotropic one in the Glauber model compared to the hard sphere model.

\begin{figure}[thb]
\includegraphics[width=0.9\linewidth]{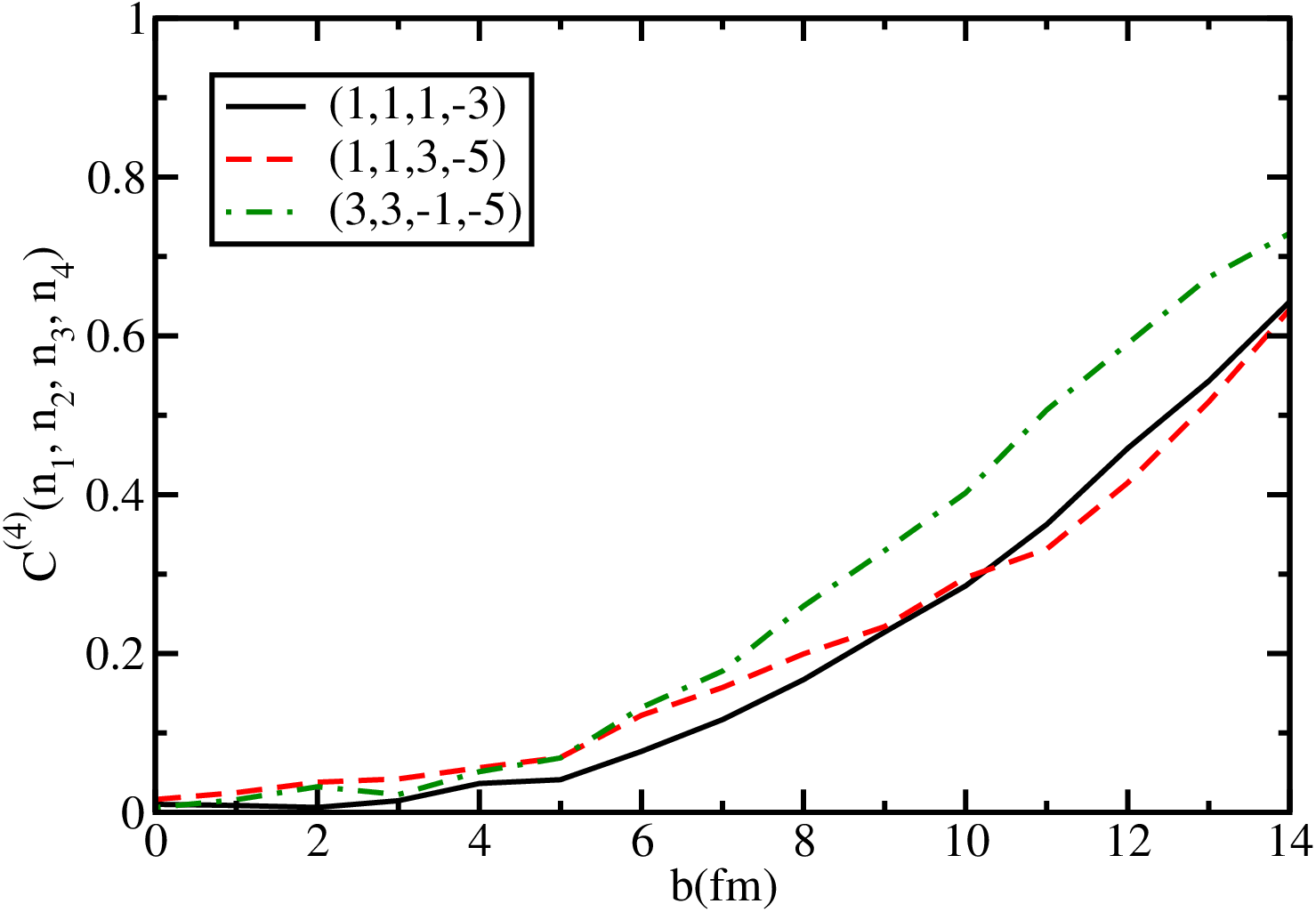}
 \caption{(Color online) The correlation functions $C^{(4)}$ from Monte-Carlo Glauber modeling of the initial conditions for Pb+Pb collisions at $\sqrt{s_{NN}} = 2.76$~TeV at the LHC.
} \label{lhc_cos}
\end{figure}

Next we calculate the correlations between different harmonic moments for realistic simulations of lead-lead collisions at the LHC energy of $\sqrt{s_{NN}} = 2.76$~TeV using the Monte-Carlo Glauber model following Ref. \cite{Qin:2010pf}, where the fluctuations of nucleon positions inside two colliding nuclei as well as the fluctuations arising from individual nucleon-nucleon collisions are included.
The results for the correlation functions $C^{(4)}$ are plotted as a function of impact parameter $b$ in Fig.~\ref{lhc_cos}.
Here we take the initial condition with free-streaming of $0.25$~fm to account for the evolution of hot spots during the pre-equilibrium stage.
One can see that the correlations between different harmonic moments increase from central collisions to peripheral collisions, indicating a decrease in the number of hot spots.
Compared with the results from Fig.~\ref{cosmn}, we obtain just a few hot spots in very peripheral collisions and up to about one hundred hot spots in most central collisions. 
We note that the number of hot spots may be different in other initial condition models, such as the Color Glass Condensate model \cite{Dumitru:2008wn, Dusling:2009ar, Gavin:2011gr}, where the number of fluctuations is essentially determined by the transverse size of the color flux tubes and the overlap area of the colliding nuclei.

In summary, we have studied the effect of local fluctuations in the initial states on the geometric anisotropy by considering the correlations between different harmonic moments. We find that the strength of such correlations strongly depends on the number of hot spots produced at initial times. Such correlations in the initial states can be directly related to the correlations in the final states owing to the hydrodynamic response to the pressure gradient of the fireball. Thus the measurement of the correlations between four different harmonic moments, if experimentally feasible, can determine how many hot spots are inside the quark-gluon plasma created in relativistic heavy-ion collisions.

%\section{Acknowledgments}
We thank C.~E.~Coleman-Smith for discussions. This work was supported in part by U.S. Department of Energy grant DE-FG02-05ER41367.

%%%%%%%%%%%%%%%%%%%%%%%%%%%%%%%%%%%%%%%%%%%%%%%%%%%%%%%%%%%%%%%%%%%%%%%%%%%%%%%%
%%%%%%%%%%%%%%%%%%%%%%%%%%%%%%%%%%%%%%%%

%%%%%%%%%%%%%%%%%%%%%%%%%%%%%%%%%%%%%%%%%%%%%%%%%%%%%%%%%%%%%%%%%%%%%%%%%%%%%%%%
%%%%%%%%%%%%%%%%%%%%%%%%%%%%%%%%%%%%%%%%%
%\bibHeading{References}
%\bibliographystyle{plain}
\bibliographystyle{h-physrev5.bst}
\bibliography{GYQ_refs.bib}
%%%%%%%%%%%%%%%%%%%%%%%%%%%%%%%%%%%%%%%%%%%%%%%%%%%%%%%%%%%%%%%%%%%%%%%%%%%%%%%%
%%%%%%%%%%%%%%%%%%%%%%%%%%%%%%%%%%%%%%%%
\end{document}